\documentclass[twocolumn,showpacs,preprintnumbers,amsmath,amssymb]{revtex4}

\usepackage{graphicx}
\usepackage{dcolumn}
\usepackage{bm}

\newcommand{\ybco}{YBa$_2$Cu$_3$O$_{7-\delta}$ }

\begin{document}


\title{Order-disorder transition induced by deformation of vortex lines at the twin boundaries in \ybco-crystals: test of the Lindemann criteria}

\author{Yu. T. Petrusenko}
\affiliation{National Science Center "Kharkov Institute of Physics and Technology", 1 Akademicheskaya St., Kharkov 61108, Ukraine
}%

\author{A. V. Bondarenko}
 \email{Aleksandr.V.Bondarenko@univer.kharkov.ua}
\author{A. A. Zavgorodniy}
\author{M. A. Obolenskii}
\author{V. I. Beletskii}
\affiliation{Physical department, V.N. Karazin Kharkov National University, 4 Svoboda Square, 61077 Kharkov, Ukraine}

\date{\today}

\begin{abstract}
We show that rotation of the magnetic field off the plane of twin
boundaries (TB's) induces transition of an ordered vortex solid
phase to a disordered one. This transition arises due to
appearance of transverse deformations of vortex lines near the

criteria, $u_{t,rp}=c_La_0$. This order-disorder transition is
accompanied by an increase in the depinning current, by crossover
from an elastic creep to plastic one, and by appearance of the
$S$-shaped voltage-current characteristics.
\end{abstract}

\maketitle

The non-monotonous field variation of the pinning force $F_p$ in
low-$T_c$ (NbSe$_2$ \cite{Bhattacharya93, Higgins96}, V$_3$Si
\cite{Gapud03}) and high-$T_c$ (BiSrCaCuO \cite{Khaikovich96},
YBaCuO \cite{Kupfer98,Pissas00}) superconductors is subject of
long-time interest. Increase of the pinning force can be explained
by softening of the elastic moduli of vortex lattice in vicinity
of the upper critical field $H_{c2}(T)$ \cite{Higgins96} or the
melting line $H_m(T)$ \cite{Kwok94} that causes better adaptation
of the vortex lines to the pinning landscape. In frames of the
collective pinning theory \cite{Blatter94} the non-monotonous
field variation of the force $F_p$ in the thermally activated mode
is determined by competition between increase of the activation
energy $U$ and decrease of the depinning current $J_d$ upon
increase of the field. Two alternative models \cite{Ertas96,
Rosenstein07} suggest transition of an ordered VL into a
disordered one with increased magnetic field, though the nature of
the order-disorder (OD) transition and increase of the force $F_p$
in these models are different. These models are supported by
correlation between the field corresponded to the structural OD
transition \cite{Cubbit93} and the onset of the $F_p$ increase
\cite{Khaikovich96} in BiCaSrCuO crystals.

\begin{figure}
\includegraphics[clip=true,width=3.2in]{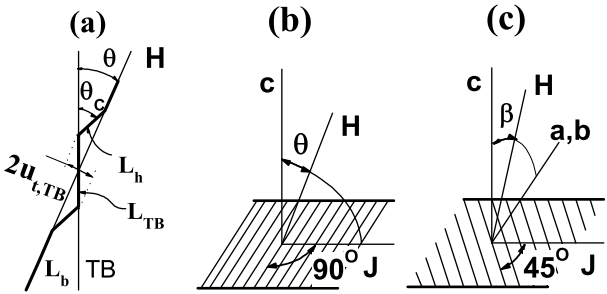}
\caption{\label{fig:1}Kinked structure of vortex line near the
plane of TB proposed in Ref.~\onlinecite{Kwok96} (panel a), and
sketch of measurements geometry of crystal C1 (panel b) and C2
(panel c).}
\end{figure}

The model proposed in Ref. ~\onlinecite{Ertas97} assumes that OD
transition occurs when transverse deformations of vortex lines
$u_{t,rp}$, induced by interaction of vortices with random pinning
potential, satisfy the Lindemann criteria, $u_{t,rp}=c_La_0$,
where $a_0\simeq(\Phi_0/B)^{1/2}$ is the intervortex distance,
$\Phi_0$ is the flux quantum, and $c_L$ is the Lindemann number.
This model can be tested in the \ybco crystals through
investigation of the effect of the magnetic field rotation off the
TB's plane on pinning and dynamics of the vortex solid (VS).
Indeed, decoration experiments \cite{Vinnikov88} show that the
superconducting order parameter at the TB's is suppressed. This
causes deformation of the vortex lines near the TB's, as it is
shown in Fig. ~\ref{fig:1}a \cite{Kwok96}. Here, at the angles
$\theta\equiv\angle\textbf{H}$,TB's smaller than a certain
critical $\theta_c$ value, a some part of the vortex line $L_{TB}$
is trapped by the TB, the vortex fragment $L_h$ and the twin plane
limit the angle $\theta_c$, and far away from the TB the vortex
line is aligned along the external field. These deformations
induce appearance of transverse displacements of the vortex line,
the amplitude of which $u_{t,TB}$ can satisfy the Lindemann
criteria in high magnetic field. Therefore, according to
Ref.~\onlinecite{Ertas97}, one can expect occurrence of the OD
transition when rotating the field off the TB's plane. Results of
our measurements give strong experimental support for occurrence
of this transition.

The measurements were performed on two \ybco crystals with $T_c
\simeq$ 93 K and $\delta T_c \simeq$ 0.4 K, which contained TB's
aligned in one direction. The transport current was applied along
the $ab$-plane and at angle of 90$^\circ$ and 45$^\circ$ to the
TB's plane in crystals C1 and C2, respectively. The average
distance between the TB's was about 0.8 $\mu$m in sample C1, and
about 0.3 $\mu$m in sample C2. Measurements were performed for
different orientations of vector \textbf{H} with respect to the
$c$-axis. In sample C1 the vector  \textbf{H} was located in plane
constituted by the vectors  \textbf{c} and  \textbf{J}, while in
sample C2 it was located in plane perpendicular to the vector
\textbf{J}, as it is shown in panel (a) and (b) of
Fig.~\ref{fig:1}, respectively. The resolution of angles $\theta$
and $\beta \equiv\angle\textbf{H,c}$ was about 0.1$^\circ$. The
concentration of point defects $n_{pd}$ in sample C2 was varied by
low-temperature ($T \leq$ 10 K) irradiation with 2.5 MeV
electrons, and the irradiation-measurements cycles were performed
without heating the sample above 110 K \cite{Bondarenko01}. The
vortex dynamics was studied through measurements of
current-voltage characteristics $E(J)$ at $dc$-current by the
standart four-probe method at a temperature of $t \equiv T/T_c$ =
0.92 in a magnetic field of 15 kOe.

\begin{figure}
\includegraphics[clip=true,width=3.2in]{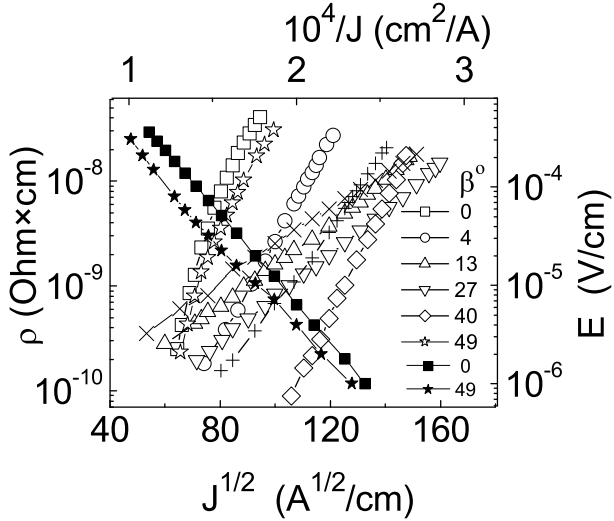}
\caption{\label{fig:2}The $E(J)$ curves in crystal C2, which are
plotted in the scale $log\rho=log[E(J)/J]$ vs. $\sqrt{J}$ (light
symbols, left-hand and bottom scales), and in the scale $logE(J)$
vs. 1/$J$ (dark symbols, right-hand and top scales). Cross-wises
symbols correspond to voltage-current characteristics plotted in
the scale $log\rho$ vs. $\sqrt{J}$, which were measured in the
parallel field after electron irradiation with doses of 10$^{18}$
(+) and 3$\cdot$10$^{18}$ ($\times$) el/cm2. Irradiation dose
10$^{18}$ produces the averaged over all sublattices concentration
of the defects 10$^{-4}$ dpa.}
\end{figure}

The measurement results of sample C2, plotted as log$E(J)$ versus
1/$J$ and log$\rho(J) \equiv$ log$E(J)/J$ versus $\sqrt{J}$, are
shown in Fig.~\ref{fig:2}. The dynamic resistance $\rho _d(J)
\equiv dE(J)/dJ$ corresponded to the measured $E(J)$ dependencies
is much smaller than the flux flow resistance $\rho _{BS}$
indicating that measurements correspond to the thermally activated
creep mode. It is seen that in non irradiated sample and at angle
$\beta$ of $\beta$ = 0$^\circ$ and $\beta$ = 49$^\circ$ the
experimental data follows equation
\begin{eqnarray}
E(J)=e_0exp[(-U_{el}/k_BT)(J_d/J)]
\label{eq:e}
\end{eqnarray}
correspondent to the elastic mechanism of vortex creep
\cite{Blatter94}, while in the interval of angles 4$^\circ \leq
\beta$ < 49$^\circ$ they follows equation
\begin{eqnarray}
E(J)=\rho_0Jexp\{(-U_{pl}/k_BT)[1-(J/J_d)^{0.5}]\}
\label{eq:e05}
\end{eqnarray}
which corresponds to the plastic creep mediated by motion of the
VS dislocations. Here $E_0$ and $\rho_0$ are the constants, and
$U_{el}$ and $U_{pl}$ are the activation energies correspondent to
elastic and plastic creep, respectively. It is also seen that in
irradiated sample the creep follows Eq.~\ref{eq:e05} at angle
$\beta$ of $\beta$ = 0$^\circ$. The crossover from elastic to
plastic creep, realized in the field
$\textbf{H}\parallel\textbf{c}$ with an increased concentration
$n_d$, is caused by OD transition \cite{Bondarenko01}. The
transition arises due to increase in the pinning energy, $E_p
\propto n_d^{1/3}$ \cite{Blatter94}, which dominates over increase
of the elastic energy induced by transverse deformations
$u_{t,rp}=c_La_0$. The crossover from elastic to plastic creep,
which is observed in a non irradiated sample at field rotation off
the TB's plane through $\beta \geq$ 4$^\circ$, can arise due to
the OD transition, too. But in this case it is initiated by
appearance of transverse displacements of vortex lines near the
TB's, $u_{t,TB}$, see Fig.~\ref{fig:1}b. The displacement
amplitude
\begin{eqnarray}
u_{t,TB}\simeq L_h\sin(\theta_c - \theta) \label{eq:uttb}
\end{eqnarray}
is specified by the length of fragment $L_h \simeq (\varepsilon
a_0/2\sqrt{\pi})[ln(a_0/\xi)]^{1/2}$ \cite{Kwok96}, and the angles
$\beta$ and $\theta$ in sample C2 are related by
$\sqrt{2}$sin$\theta$ = sin$\beta$. Decoration experiments
\cite{Herbsommer00} show that TB's affect vortex structure up to
angle $\theta_c \simeq$ 70$^\circ$, when the field is rotated off
the $c$-axis. For $a_0$(15kOe) = 400~\AA, $\theta$ = 4$^\circ$,
$\theta_c \simeq$ 70$^\circ$, and for reasonable values of the
anisotropy parameter $\varepsilon$ = 1/5 and coherence length
$\xi$ (85K) = 40~\AA~ we obtain the amplitude $u_t \simeq$
0.1$a_0$, which satisfy the Lindemann criteria.

\begin{figure}
\includegraphics[clip=true,width=3.2in]{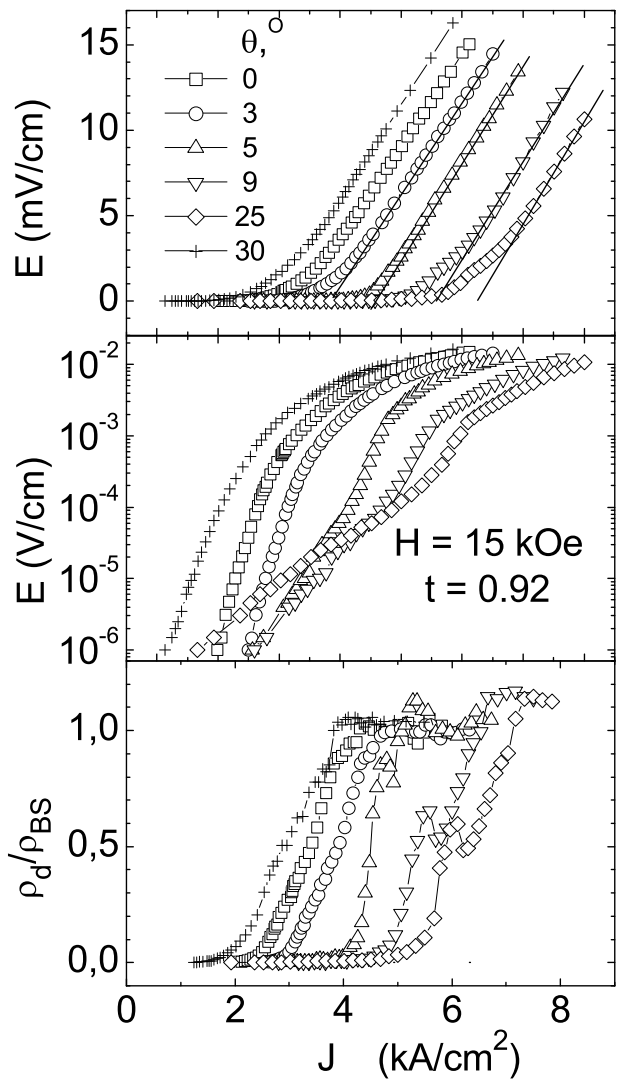}
\caption{\label{fig:3} The $E(J)$ curves in crystal C1 plotted in
the linear (panel a) and semi-logarithmical (panel b) scale. Panel
c  shows the $\rho_d(J)/\rho_{BS}$ curves plotted in the linear
scale.}
\end{figure}

\begin{figure}
\includegraphics[clip=true,width=3.2in]{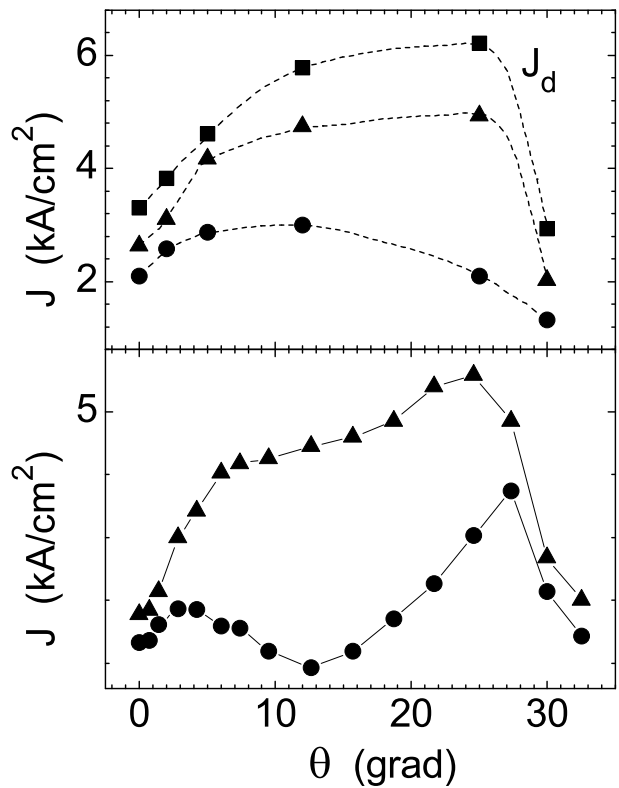}
\caption{\label{fig:4} Angular variation of the depinning current
$J_d$ (squares) and of the currents $J_{E1}$ and $J_{E2}$
determined at electric field level of $E_1$ = 2 $\mu$V/cm
(circles) and $E_2$ =100 $\mu$V/cm (triangles), respectively, in
crystal C1 (panel a) and C2 (panel b).}
\end{figure}

The crossover from elastic to plastic creep is observed in the
sample C1 too. In this sample, as evident from Fig.~\ref{fig:3}c,
the value of ratio $\rho_d/\rho_{BS} \ll$1 corresponds to the
creep regime at small currents, while at high currents the ratio
$\rho_d/\rho_{BS} \simeq$1 corresponds to the flux flow regime.
Angular variation of the critical currents $J_{E1}$ and $J_{E2}$
inside the creep regime (determined at voltage criteria of 2 and
100 $\mu$V/cm, respectively) and of the depinning current $J_d$
(determined by extrapolation of the linear parts of the $E-J$
curves in zero voltage) is shown in Fig.~\ref{fig:4}a. The
currents $J_d$ and $J_{E1}$ vary with the angle $\theta$ in a
similar way: they gradually increase with angle $\theta$ up to the
value of $\theta$=25$^\circ$, and then sharp drop down to their
values at angle $\theta$ = 0$^\circ$. Fig.~\ref{fig:4}b shows
angular variation of the critical currents $J_{E1}$ and $J_{E2}$,
determined at the same voltage criteria, in sample C2.

The rise in $J_c$, observed in the angular range 0 $< \theta \leq$
25$^\circ$ can not be explained by enhanced pinning of the trapped
fragments $L_{TB}$ because their portion (the value of ratio
$L_{TB}/L_b$) decreases as sin($\theta_c-\theta$). Also, the
reduced pinning in the field $\textbf{H}\parallel\textbf{c}$ can
not be explained by suppression of pinning by point defects, as it
was assumed in Ref.~\onlinecite{Solovjov94}: the vortices located
in between the TB's poorly accommodate themselves to the point
defects landscape due to a strong interaction with vortices
trapped by the TB's. This interpretation implies the formation of
the ordered VS in the irradiated samples placed in the parallel
field; that contradicts the results of our measurements. Therefore
we believe that the current $J_d$ increases due to occurrence of
the OD transition. This interpretation implies that weak
1D-pinning of the ordered VS, which realizes in the parallel
field, is replaced by strong 3D-pinning of the disordered VS in
the inclined fields. Density of the displacements (number of
displacements $n_{t,TB}$ per unit vortex length) increases as
$n_{t,TB} \propto sin\theta$, this leading to continuous increase
of the dislocation concentration, and thus, to a greater disorder
of the vortex solid. Therefore the current $J_d$ increases due to
better adaptation of the disordered phases to the pinning
landscape \cite{Gingras96} and \cite{Giamarchi97}. On the other
hand, according to Eq.~\ref{eq:uttb} the amplitude $u_{t,TB}$
decreases with increased angle $\theta$. Therefore, as soon as it
drops below the value of $c_La_0$, one can expect transformation
of the disordered VS into the ordered VS. This transition must be
accompanied by the crossover of the 3D to 1D pinning regime that
reduces the pinning force, and by the crossover of the plastic to
elastic creep regime. Sharp drop in the current $J_c$ and
crossover from the plastic to elastic mechanism of vortex creep,
which are realized in the narrow interval of angles 25$^\circ
\leq\beta\leq$ 30$^\circ$, strongly support occurrence of this
transition.

An important feature of vortex dynamics is that the resistance
$\rho_d$ monotonous increases with the current at angles $\theta$=
0$^\circ$ and $\theta$ = 30$^\circ$, while in the interval of
angles 5$^\circ\leq\beta\leq$ 25$^\circ$ a peak in the $\rho_d(J)$
curves is observed, see Fig.~\ref{fig:3}c. This peak corresponds
to the $S$-shape of the $E(J)$ curves, which in some papers is
attributed to possible non homogeneous distribution of the
magnetic flux or current in samples. However, it is difficult to
point out any reliable mechanism responsible for the change of
homogeneity of these parameters with small change in the angle
$\theta$, namely, with it variation from 0 to 5$^\circ$, or from
25$^\circ$ to 30$^\circ$. Therefore we believe that the peak is
dynamic characteristic of the disordered VS, and it characterizes
the dynamic ordering of the VS in presence of the random pinning
potential, as it is observed in the numerical studies
\cite{Faleski96,Kolton99}. This ordering is caused by reduction in
the amplitude $u_t\propto 1/v$ \cite{Koshelev94}, and realizes in
the interval of currents confined by the peak and minimum position
in the $\rho_d(J)$ curves \cite{Faleski96,Kolton99}. The effect of
dynamic ordering on the pinning of VS in \ybco crystals containing
chaotic pinning potential has been recently studied in
Ref.~\onlinecite{Bondarenko08b}. It has been found that dynamic
ordering substantially reduces the pinning force. Note that in
crystal C1 vortices move along the plane of TB's and the amplitude
$u_{t,TB}$ is not changed with $v$ due to 2D nature of these
defects. Therefore increase in $v$ partially orders the VS, but
dynamic VS remains disordered due to permanent amplitude
$u_{t,TB}$ that causes substantial increase of the current $J_d$
in the interval of angles 0 $< \theta\leq$ 25$^\circ$.

As seen in Fig.~\ref{fig:4}a, angular variation of the current
$J_{E1}$, which characterizes pinning in deep creep regime,
differs from the $J_d(\theta)$ and $J_{E2}(\theta)$ dependencies.
This is reasonable considering that inside deep creep regime the
pinning force depends on both the depinning current and activation
energy. The activation energy correspondent to the plastic creep
depends on the angle $\alpha\equiv\angle\textbf{H},ab$
\cite{Bondarenko01} and amplitude $u_{t,TB}$ \cite{Bondarenko01b}.
It can be shown that in presence of correlated displacements
$u_{t,TB}$ the energy $U_{pl}$ depends on mutual orientation of
the displacements and direction of vortex motion. This is caused
different angular variation of the current $J_{E1}$ in samples C1
and C2. Detailed analysis of this difference and of the effect of
point disorder on angular variation of the energy $U_{pl}$ and on
angular variation of the currents $J_{E1}$ and $J_{E2}$ will be
discussed elsewhere \cite{Bondarenko08c}.

In conclusion, we have studied the effect of transverse
deformation of vortex lines near the planes of TB's on the pinning
force, mechanism of thermally activated creep, and dynamics of
vortex solid. We show that in the interval of angles 5
$\leq\theta\leq$ 25$^\circ$ the amplitude of displacements
$u_{t,TB}$ satisfy the Lindemann criteria, $u_{t,TB}= c_La_0$,
that leading to formation of the disordered vortex solid. This
phase is characterized by the plastic creep mediated by motion of
dislocations, by increase of the depinning current with increased
density of the displacements, and by the $S$-shaped
voltage-current characteristics, which manifests partial dynamic
ordering of the vortex solid induced by suppression of the effect
of random point pinning potential. Decrease of the amplitude
$u_{t,TB}$ below the value of $c_La_0$ causes transition to the
ordered vortex solid, which is characterized by the elastic creep
and smaller depinning current.


\bibliography{/bondarenko/tex.sample/paper}

\end{document}